\documentclass[twocolumn,floats,floatfix,superscriptaddress,prx,nofootinbib]{revtex4-1}
\usepackage{bm}
\usepackage{graphicx}
\usepackage{amsmath}
\usepackage{amsfonts}
\usepackage{mathrsfs}
\usepackage{bbold}
\usepackage{hyperref}
\usepackage[utf8]{inputenc}

\begin{document}

\title{The Geometry of the Cholesteric Phase}

\author{Daniel A. Beller}
\affiliation{Department of Physics and Astronomy, University of Pennsylvania, Philadelphia, PA 19104-6396, USA}
\author{Thomas Machon}
\affiliation{Department of Physics and Centre for Complexity Science, University of Warwick, Coventry CV4\,7AL, UK}
\author{Simon \v{C}opar}
\affiliation{Department of Physics and Astronomy, University of Pennsylvania, Philadelphia, PA 19104-6396, USA}
\affiliation{Faculty of Mathematics and Physics, University of Ljubljana, Jadranska 19, 1000 Ljubljana, Slovenia}
\affiliation{Jozef Stefan Institute, Jamova 39, 1000 Ljubljana, Slovenia}
\author{Daniel M. Sussman}
\affiliation{Department of Physics and Astronomy, University of Pennsylvania, Philadelphia, PA 19104-6396, USA}
\author{Gareth P. Alexander}
\affiliation{Department of Physics and Centre for Complexity Science, University of Warwick, Coventry CV4\,7AL, UK}
\author{Randall D. Kamien}
\affiliation{Department of Physics and Astronomy, University of Pennsylvania, Philadelphia, PA 19104-6396, USA}
\author{Ricardo A. Mosna}
\affiliation{Department of Physics and Astronomy, University of Pennsylvania, Philadelphia, PA 19104-6396, USA}
\affiliation{Departamento de Matem\'atica Aplicada, Universidade Estadual de Campinas, 13083-859, Campinas, SP, Brazil}

\date{\today}

\begin{abstract}
We propose a construction of a cholesteric pitch axis for an arbitrary nematic director field as an eigenvalue problem.  Our definition leads to a Frenet-Serret description of an orthonormal triad determined by this axis, the director, and the mutually perpendicular direction.  With this tool we are able to compare defect structures in cholesterics, biaxial nematics, and smectics.  Though they all have similar ground state manifolds, the defect structures are different and cannot be, in general, translated from one phase to the other.  
\end{abstract}
\newcommand\nbar{{\bf/\mkern-11mu n}}
\newcommand\notn{/\mkern-11mu n}
\newcommand{\Tr}{\mathop{\rm Tr}}
\renewcommand\vec[1]{\mathbf{#1}}

\maketitle
The standard lore in liquid crystal physics is that the cholesteric and the smectic are essentially the same beast: the Caille instability in smectics \cite{caille}\ is the same as the Lubensky instability in cholesterics \cite{lubensky}, both examples of the Peierls instability.  Fluctuations around the ground states of each are dominated by a single Goldstone mode with an unusual fluctuation spectrum that distinguishes modes along the layer normal or pitch axis from modes along other directions.   Since these modes are also the co\"ordinates on the ground state manifold (GSM), one is tempted to identify defects in smectics with defects in cholesterics, with the cholesteric pitch axis $\bf P$ analogous to the smectic layer normal $\bf N$.  

On the other hand, the cholesteric phase is biaxial \cite{biaxchol,hkl}: intuitively, a cholesteric is defined by a unit director field $\bf n$ and a {derived} pitch axis $\bf P$ that describes the direction along which $\bf n$ twists, whose precise definition we will discuss below. At each point, a local orthonormal triad is specified by $\bf n$ and $\bf P$, both defined up to sign, along with their cross product $ \bf n \times \bf P \equiv {\bf \nbar} $ (``not $\bf n$'').  From this point of view, one might attempt to identify defects in the cholesteric with defects in the biaxial phase characterized by the standard homotopy theory of topological defects \cite{mermin,monas,vm,klemanchol2}, in particular via the non-Abelian fundamental group $\pi_1[SO(3)/{\cal D}_{2d}]= \{\pm \mathbb{1},\pm \sigma_x,\pm\sigma_y,\pm\sigma_z\}$, the unit quaternions.  This has been the motivation behind a classification of cholesteric defects as singularities in two of three triad directions, termed the $\lambda^{\pm 1}$, $\tau^{\pm 1}$, and $\chi^{\pm 1}$ singularities, in correspondence with $\pm \sigma_x$, $\pm \sigma_y$, and $\pm \sigma_z$ singularities of the biaxial nematic phase \cite{klemanchol}. For example, just as $\sigma_x \sigma_y = \sigma_z$ in the quaternion group, composing a $\lambda$ with a $\tau$ yields a $\chi$ in the cholesteric. Meanwhile, the cholesteric-smectic analogy suggests that disclinations in the layer normal $\bf N$ are to be identified with the $\lambda$ and $\tau$ defects where $\bf P$ {is not defined}, so that the decomposition of a smectic dislocation into a disclination pair \cite{cak} is also analogized to the decomposition of a $\chi$ into a $\lambda$-$\tau$ pair.

In order to explore the similarities and differences between the topology of defects in smectics, cholesterics, and biaxial nematics, we will develop the geometric tools necessary to map between these three systems.  We find that though there is a natural mapping between the ground states of these three systems, the descriptions of their defects are all different.  In particular, we demonstrate the failure of this natural mapping in three example textures: the $\chi^2$ to $\lambda^2$ defect transformation, the oblique helicoidal cholesteric texture, and the screw dislocation.  Our result highlights the well-known problem with classifying topological defects in translationally ordered media \cite{mermin,poenaru,cak} and confronts  the usual analogies between our triptych of phases.  

There are crucial differences between the three systems, illustrated most clearly in the symmetries of their ground states. A smectic ground state with layer normal $\hat z$ and layer spacing $a$ is invariant under a translation by $a$ along $\hat z$; in other words, a continuum of ground states is available under translations modulo $a$ along $\hat z$. This discrete symmetry is associated with the existence of dislocations in the smectic phase. The smectic ground state also possesses a continuous symmetry under arbitrary rotations about $\hat z$. In contrast, the biaxial nematic ground state with director $\bf n$ parallel to $\hat z$ possesses a continuous symmetry under arbitrary translations along $\hat z$, but a discrete symmetry under rotations by $\pi$ about $\hat z$. The latter discrete symmetry is related to the biaxial singularities not found in uniaxial nematics.  Along with uniaxial nematics, smectics and biaxial nematics are invariant under $\pi$ rotations around $\hat x$ and $\hat y$.

From the point of view of symmetries the cholesteric is intermediate between the smectic and biaxial nematic phases. A cholesteric ground state with pitch axis  $\hat z$ and helical wave number $q_0$ has a continuous screw symmetry: rotations by an angle $\alpha$ about $\hat z$ composed with translations by $\alpha/q_0$ along $\hat z$. However, a translation by $\pi/ q_0$ along $\hat z$, and a rotation by $\pi$ about $\hat z$, are each (separately) discrete symmetry operations of the cholesteric ground state (as are rotations by $\pi$ about $\hat x$ or $\hat y$). In other words, the cholesteric ground state has the discrete symmetries of both the smectic and the biaxial nematic ground states, which might lead us to expect that cholesteric defects will bear some resemblance to both biaxial nematic defects and smectic defects. However, broken translational invariance leads to a breakdown of the homotopy theory of defects \cite{mermin,poenaru,cak} and so symmetries are not the whole story.

We note also that smectics can have point defects, for instance when the layers are concentric spheres around a single point, whereas the biaxial nematic cannot support such defects as $\pi_2[SO(3)/{\cal D}_{2d}]=0$.  This makes the interpretation of the cholesteric even more subtle.  From one point of view the cholesteric configuration is completely specified by a director field everywhere in space and is thermodynamically equivalent to a uniaxial nematic, which has ground state manifold $\mathbb{R}P^2$ and can support point defects.  However, as discussed in \cite{SVD} if we include the pitch axis then a simple point defect in the director field necessarily sprouts line defects in the pitch axis.  From this point of view the connection between the cholesteric and biaxial nematic may seem promising.  However, there is a weakness in this link as well, which  will appear when we give an explicit definition of the $\bf P$.

To proceed any further, let us now define the unit cholesteric pitch axis $\bf P$.  With the orthonormal triad of the biaxial in mind, we {require} ${\bf P}\cdot{\bf n}=0$ and define the direction ``not $\bf n$'' as  $\nbar={\bf n}\!\times\!{\bf P}$; the cholesteric triad is then $\{{\bf n}, {\bf P}, {\bf \nbar}\}$. Intuitively, $\bf P$ is a direction about which $\bf n$ twists, meaning that the directional derivative $({\bf P}\cdot \boldsymbol{\nabla}){\bf n}$ has no component along $\bf P$. As $\bf n$ is a unit vector, this directional derivative is also perpendicular to $\bf n$, so we can define $\bf P$ by the equation $(\vec{P}\cdot\boldsymbol{\nabla})\vec{n}=-q\,\nbar$, or the equivalent eigenvalue problem
\begin{equation}\label{eq:C}
P_iC_{ij} =q P_j
\end{equation}
where $q$ is the local eigenvalue and $C_{ij} = n_k\epsilon_{\ell j k}\partial_i n_\ell$ is precisely the same
chirality tensor defined in \cite{efrati}.  Note, in addition, that ${\rm Tr}\; C = -{\bf n}\cdot{\boldsymbol{\nabla}\!\times\!{\bf n}}$.  This approach is similar to, but more restricted than, the definition proposed in \cite{straley} that only holds for the cholesteric ground state texture.  Our construction {generates} a triad defined up to sign so the GSM is $SO(3)/{\cal D}_{2d}$.  

Whence the weak link in the analogy between biaxial nematics and cholesterics?  Note that in the biaxial nematic, the local triad can undergo a global rotation that may change the energy but not the topology, in precise analogy with
the two-dimensional uniaxial nematic; a $+1$
disclination, for example, can be purely bend (azimuthal) or purely splay (radial). If the
bend and splay elastic constants differ, the energies will differ but
the topology will not.  However, this internal rotation is not a symmetry of the cholesteric since the pitch axis, defined through $C_{ij}$, involves spatial derivatives.    As a result, rotations of the triad also require a rotation of the system, {\sl i.e.}\ only the trivial rotation of the whole sample leaves the topology invariant.  Though it is tempting to view the cholesteric's $\lambda$, $\tau$, and $\chi$ defects in analogy with the $\sigma_x$, $\sigma_y$, and $\sigma_z$ defects of the biaxial nematic, the correspondence is not precise: Homotopies between biaxial nematic triad configurations, which establish the topological equivalence of two configurations, are not in general available to cholesteric triads, where $\bf P$ is determined by spatial derivatives of $\bf n$ and cannot rotate freely.

\begin{figure}[t] 
\includegraphics[width=\linewidth]{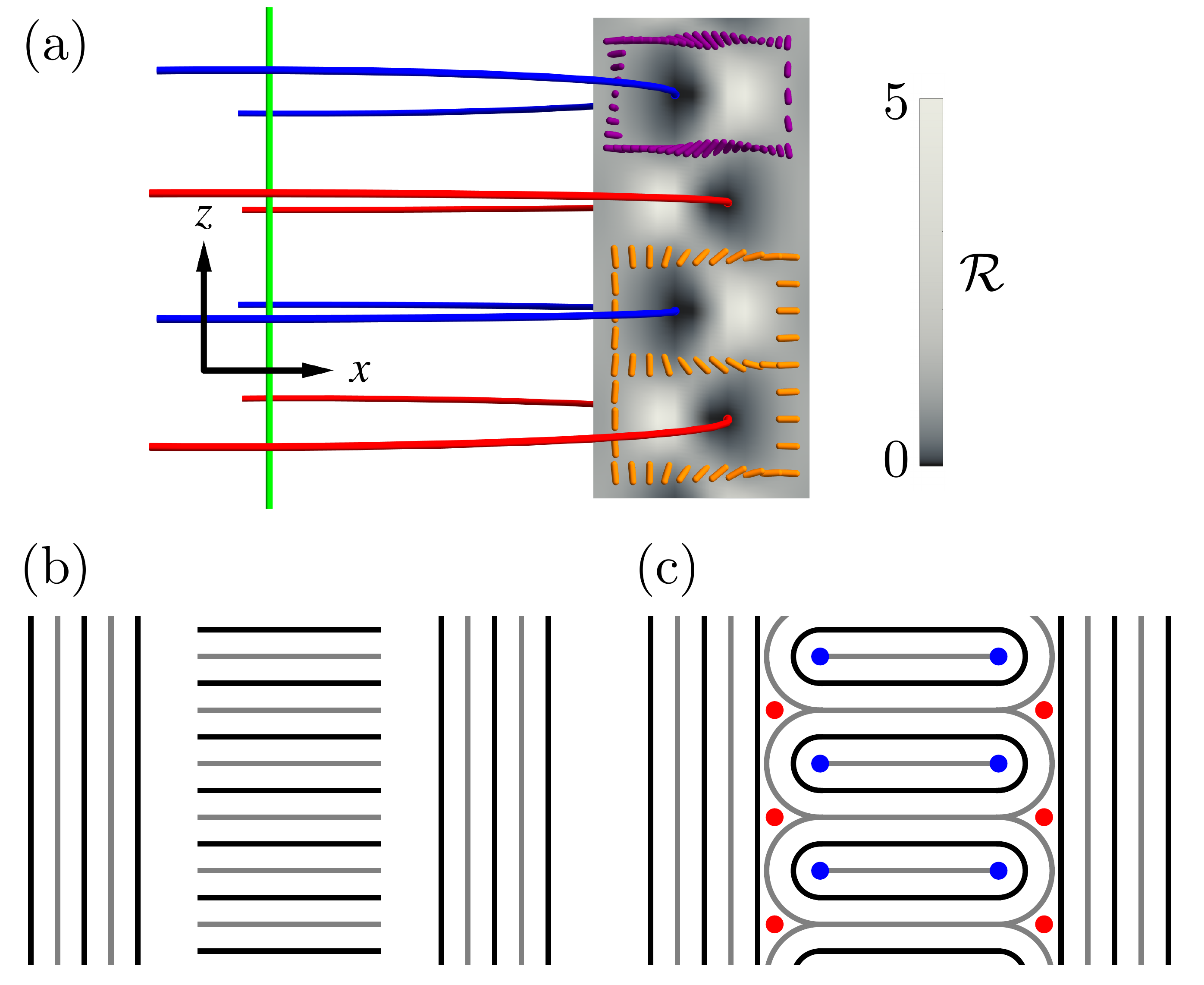}
\caption{The cholesteric interpolation of Equation (\ref{interp}).  (a) A $\chi^2$ line (green) at $r=0$ surrounded by $\lambda$ (blue) and $\lambda^{-1}$ (red) rings near $r=r_0$. The grey scale is a density plot of ${\cal R}$ in a portion of the $xz$ plane; zeros (black) indicate defects in $\vec{P}$. Orange rods show $\vec{P}$ and its nontrivial winding around the defect rings; purple rods show $\vec{n}$, which is nonsingular. (b) Level sets of the cholesteric $\phi$ field, in the small and large $r$ limits. (c) Filling space between the two boundaries requires dislocations in the layers, realized as $\pm \pi$ disclination pairs.  $\phi \in \mathbb{Z}$ are in black; half-layers are in gray. \label{interpfig} }
\end{figure}

As a result, it is not possible to fully preserve the algebraic topology of the biaxial triad in the cholesteric.  For example, in the biaxial nematic $\sigma_i^2=-\mathbb{1}$ for all $i=x,y,z$.  This implies that there is a single homotopy class for defects with $2\pi$ winding, independent of the triad direction around which the rotation is performed. Thus, a pair of $\sigma_x$ singularities could be brought together and, through a continuous family of biaxial nematic configurations, could be transformed into a pair of $\sigma_z$ singularities. However, this equivalence spectacularly fails in the cholesteric.  Let us see what happens when we try to transform a $\chi^2$ defect configuration, with $2\pi$ rotation about $\bf P$,
${\bf n}_{\chi^2} =\hat r\sin(q_0 z) + \hat\phi\cos(q_0 z)$ (in cylindrical co\"ordinates),
into a $\lambda^2$ defect configuration, with $2\pi$ rotation about $\bf n$,
${\bf n}_{\lambda^2} =  \hat z \cos(q_0 r) + \hat\phi\sin(q_0 r)$.
In the $\bf n_{\chi^2}$ configuration, $\bf P$ is vertical and $\bf n$ is in a planar $+1$ disclination configuration, rotating between the radial and azimuthal geometries as a function of $z$. The ${\bf n}_{\lambda^2}$ configuration is the classic double-twist configuration, with a radial $\bf P$. Consider the interpolation
\begin{equation}
{\bf n} =\frac{{\bf n}_{\chi^2} \cos\Theta(r)+ {\bf n}_{\lambda^2}\sin \Theta(r)}{\sqrt{1+{\bf n}_{\chi^2}\cdot{\bf n}_{\lambda^2}\sin2\Theta(r)}}
\label{interp}
\end{equation}
where $\Theta(r)$ is a sigmoidal function ranging from $0$ to $\pi/2$.  To be concrete we pick
$\Theta(r) =\pi\left[ (1-e^{-r})/(1+e^{-(r-r_0)})\right]/2$
with $r_0=4\pi/q_0$.  We discover that this transformation from ${\bf n}_{\chi^2}$ to ${\bf n}_{\lambda^2}$ gives rise to a series of $\lambda^{\pm 1}$ ring defects, with alternating winding sign, encircling the central $\chi^2$ defect at radii close to $r_0$ (Fig.~\ref{interpfig}a), an iconic structure considered long ago \cite{bkold}.
Thus the straightforward homotopy between $\sigma_x^2$ and $\sigma_z^2$ defects in the biaxial nematic does {\sl not} carry over to a homotopy between the supposedly analogous $\lambda^2$ and $\chi^2$ defect configurations. Instead, the transformation is mediated by the appearance of another set of defects. We will see below that these $\lambda^{\pm 1}$ ring defects are made necessary by  smectic-like features of the cholesteric phase.

Just as the analogy between cholesteric and biaxial nematic ground states leads us astray in the study of defects, the intuitive correspondence between cholesteric and smectic ground states fails when extended to general configurations. We can make a straightforward identification of the cholesteric pitch axis $\bf P$ with a smectic layer normal $\bf N$. With ${\bf P} = \hat z$, we could draw a level set wherever the cholesteric's director $\bf n$ points in a particular direction, say $\hat x$, and identify these level sets with smectic layers. Indeed, when we view a cholesteric with $\bf P$ in the plane of the image through crossed polarizers we see alternating bright and dark stripes; when the cholesteric has distortions we may see a distorted pattern of stripes, as in the famous ``fingerprint'' patterns. The problem with thinking of cholesterics as a ``layered'' phase, however, is that the pitch axis need not be normal to {\sl any} set of layers. Nothing requires the pitch axis to satisfy the Frobenius integrability condition {\cite{Aminov}}, ${\bf P}\cdot {\boldsymbol{\nabla}} \times {\bf P}=0$, which is necessary for $\bf P$ to be normal to a family of surfaces defined by level sets of some phase field $\phi(\bf x)$.

As an example, consider the recently observed oblique helicoidal cholesteric phase \cite{Xiang2014electrooptic}, with director field ${\bf n}=[\cos\beta\cos kz,\cos\beta\sin kz,\sin\beta]$ for some tilt angle $\beta$.  Using our construction we find that ${\bf P}=[-\sin\beta\cos kz,-\sin\beta\sin kz,\cos\beta]$ with eigenvalue $q=k\cos^2\beta$. Moreover, ${\bf P}\cdot(\boldsymbol{\nabla}\!\times\!{\bf P})=-k\sin^2\beta$ and so this conical texture, existing as an excited cholesteric state, cannot be foliated, so there is no corresponding smectic structure whose layer normal $\bf N$ agrees with $\bf P$.  There is, however, a linear combination of $\vec{P}$ and $\vec{\nbar}$ whose twist does vanish everywhere, providing a suitable $\boldsymbol{\nabla} \phi$ direction. We will return to this example below after discussing the physical interpretation of $\phi$ in a cholesteric. Herein lies the central difficulty in connecting the defect classification between cholesterics, smectics, and the biaxial nematic.  In order to identify cholesteric $\chi$ disclinations with smectic dislocations, it is necessary to foliate space; this is impossible with our choice of pitch axis, but our pitch axis is {\sl necessary} to establish a triad to connect to the GSM of the biaxial nematic.

Not all uniaxial nematic configurations can be described as cholesteric, for instance the achiral nematic ground state.  The breakdown of the cholesteric description of nematics is also encoded in the tensor $C_{ij}$. Because the tensor is not symmetric, its eigenvalues need not be real; when they are complex no pitch axis can be identified in the director field by our prescription {\cite{f2}}.  Consider the characteristic polynomial for $C_{ij}$
\begin{equation}
  -q^3+q^2 \Tr C-q \Gamma(C)+{\det C}=0
\end{equation}
where $2\Gamma(C)=\boldsymbol{\nabla}\cdot\left[\vec{n}\left(\boldsymbol{\nabla}\cdot\vec{n}\right) - \left(\vec{n}\cdot\boldsymbol{\nabla}\right)\vec{n}\right]$, the saddle-splay.
Since $\bf n$ is a null right eigenvector of $C_{ij}$, one eigenvalue vanishes.  Reality of the remaining two requires that ${\cal R}=(\Tr C)^2 -4 \Gamma(C)\ge0$
where we dub ${\cal R}$ the ``cholestericity'' -- when it is positive we have two distinct pitch axes \cite{bg}, when it is negative there are no real values of $q$ and there are no pitch axes, and when ${\cal R}=0$ we only have one pitch and one or two pitch directions, since we are not guaranteed a full basis when $C\ne C^{\tiny T}$.
When the saddle-splay vanishes  the two eigenvalues are $-\vec{n}\cdot\left(\boldsymbol{\nabla}\!\times\!\vec{n}\right)$ and $0$, and so the pitch axis is the vector corresponding to the only nonzero eigenvalue.  On the other hand, when ${\bf n}$ is a surface normal the twist vanishes, and so $C_{ij}$ is traceless.  In that case, whenever the surface has negative Gaussian curvature \cite{geometryrmp} there are two equal and opposite eigenvalue ``chiralities'' $q_2=-q_1$.  This is the origin of the directionally dependent chirality discussed by Efrati and Irvine \cite{efrati} in saddle surfaces.  A corollary of this is that surfaces with positive Gaussian curvature cannot have a real ``pitch eigenvalue.''  Only at points where ${\cal R}=0$ can the pitch axis degenerate to the circle of directions perpendicular to $\bf n$. Therefore, ${\cal R}$ vanishes {around} $\lambda^{\pm 1}$ and $\tau^{\pm 1}$ defect lines where $\bf P$ is ill-defined, just as the nematic degree of order $S$ vanishes {around} nematic defects where $\bf n$ is ill-defined. For example, the $\lambda^{\pm 1}$ rings that appeared above in the transition between ${\bf n}_{\chi^2}$ and ${\bf n}_{\lambda^2}$ appear where ${\cal R}=0$ \, as displayed in Fig.~\ref{interpfig}a.  

Of the two pitch axes identified by $C_{ij}$ when ${\cal R}>0$, the {\sl physical} pitch axis depends on the boundary conditions.  We assume that at infinity (or much closer) we have a standard cholesteric ground state with vanishing saddle-splay and thus one pitch axis.  As long as ${\cal R}>0$ we can follow the pitch axis at infinity into the bulk of the sample.  Regions where there is no pitch axis or with degenerate pitch axes are defects in the cholesteric.  Though allowed in principle from the structure of the eigenvalue problem, we do not know if it is physically possible for there to be regions in which ${\cal R}=0$ and there is only one pitch axis.  We leave this for future study.

Fortuitously our definition of the pitch axis, used to describe the rotation of ${\bf n}$,  specifies the rotation of the entire cholesteric triad as we move along $\bf P$. This will  have important consequences for our study of topological defects. Note that in addition to $\left({\bf P}\cdot\boldsymbol{\nabla}\right)\vec{n}=-q\,\nbar$, $\vec{P}\cdot\vec{n}=0$ and $\vec{P}^2=1$, which imply
$\left({\bf P}\cdot\boldsymbol{\nabla}\right) {\bf P}  = \kappa\nbar$
where $\kappa$ is a function of location. We can see that $\kappa$ is the curvature of the integral curves of $\vec{P}$.  In the manner of the Frenet-Serret apparatus \cite{geometryrmp}, we have:
\begin{equation}\label{eq:FS}
\left({\bf P}\cdot\boldsymbol{\nabla}\right) \left[\begin{matrix}\vec{P}\\ \nbar\\ \vec{n}\end{matrix}\right] = \left[\begin{matrix} 0 & \kappa &0\\-\kappa &0&q\\ 0&-q&0\end{matrix}\right]\left[\begin{matrix}\vec{P}\\ \nbar\\ \vec{n}\end{matrix}\right]
\end{equation}
Thus, the torsion of integral curves of $\bf P$ equals the local cholesteric wavenumber, provided that $\kappa$ does not vanish.  This approach leads to an alternate construction of the ``mean torsion'' \cite{torsion,Aminov}: the two nontrivial eigenvalues of $C_{ij}$ are the possible torsions of the two integral curves with binormal $\vec{n}$ and $\frac{1}{2}{\rm Tr}\;C$ is their average, {\sl the mean torsion}.  When  $C_{ij}$ has two eigenvalues $q_1\ne q_2$ we denote the two eigenvectors ${\bf P}^{(1)}$ and ${\bf P}^{(2)}$, respectively, and $\nbar^{(i)}\equiv {\bf n}\times{\bf P}^{(i)}$.  Because ${\bf n}\cdot{\bf P}^{(i)}=0$ we have ${\bf P}^{(2)} =  {\bf P}^{(1)} \cos\delta+ \nbar^{(1)}\sin\delta$ with ${\bf P}^{(1)}\cdot{\bf P}^{(2)}=\cos\delta$.  It follows from (\ref{eq:FS}) that $\boldsymbol{\nabla}\cdot{\bf n} =\nbar\cdot(\nbar\cdot\boldsymbol{\nabla}){\bf n}= (q_1-q_2)\cot\delta$; when there is splay the pitch axes are {\sl not} orthogonal.

Following streamlines of the pitch axis, we integrate the local $q$ to determine the total rotation of $\bf n$ about $\bf P$ between two points. In the ground state the angle made by the director at $z=z_0$ with the director at $z=0$ is simply $q z$ when ${\bf P} = \hat z$. We could define a phase field $\phi({\bf x})=q z$ for the ground state that measures the total rotation of $\bf n$ about $\bf P$ from a particular reference height such as $z=0$. For a general, distorted cholesteric, the Frenet-Serret apparatus enables us  to define a phase field $\phi({\bf x})$ generally, with the change in $\phi$ along integral curves of $\bf P$ recording the integrated rotation of $\bf n$ about $\bf P$. Suppose that a particular pitch axis integral curve ${\cal P}$ passes through a point $\bf x$ after traversing arc length $s$ from a reference point where $\phi=0$. Then the phase field is
$\phi({\bf x}) = \int_{\cal P} q({\bf x}) ds$
and ${\boldsymbol{\nabla}} \phi \cdot {\bf P} = q$.

We will show that the general cholesteric configuration does not admit a smectic phase field whose gradient is everywhere parallel to $\bf P$. However, the phase field $\phi$ can be defined locally for all cholesterics, so a layered structure can be found lurking inside any cholesteric configuration, even if the level sets do not correspond to obvious features of a micrograph or illustration.
On a simply connected domain without defects,  this phase field is also globally  defined and the phase field modulo $2\pi$ defines layers of the cholesteric texture. However, in the presence of line defects the absolute phase depends on the homotopy class of the integral curve from the reference point. Unlike the phase of the density fluctuations in smectics, the pitch streamlines need not generate holonomy in multiples of $2\pi$.
Thus, even with the phase field, a cholesteric in general cannot be decorated with continuous layers, though it is possible in the absence of line defects. In the oblique helicoidal configuration we considered above, one can find a direction ${\bf \tilde P}$ that is both perpendicular to $\bf n$ and satisfies ${\bf \tilde P} \cdot ({\boldsymbol{\nabla}}\times {\bf \tilde P})=0$, so that there exists a phase field $\phi$ whose gradient points along ${\bf \tilde P}$. Writing this direction as ${\bf \tilde P} \propto {\bf P} + f {\bf \nbar}$, we are led to a first order Ricatti equation for $f$ that can be solved in smooth regions of the cholesteric.  Though $\bf\tilde P$ can be foliated, we know of no general argument that relates ${\bf\tilde P}$ to $\boldsymbol{\nabla}\phi/\vert\boldsymbol{\nabla}\phi\vert$.

Do dislocations in a smectic correspond to $\chi$ defects in cholesterics?  Note that the ground states of the cholesteric and smectic share a discrete symmetry under translations by $\pi/q$ along the $\bf P$ direction. This discrete symmetry makes possible the dislocation defects suffered by smectics. Thus we might expect dislocation defects in the level sets of $\phi$ in cholesterics, as well. As in smectics, we would expect such dislocations in $\phi$ to be realized as disclination dipoles, with $\pm \pi$ winding of $\boldsymbol{\nabla} \phi$. If the eigenvalue $q$ is not allowed to vanish except at singular lines, then a nontrivial winding in $\boldsymbol{\nabla} \phi$ implies a nontrivial winding in $\bf P$, since the dot product of the two vectors never changes sign. Therefore, defects in $\boldsymbol{\nabla} \phi$ are also defects in $\bf P$, either $\lambda$ or $\tau$ singularities.  Prohibiting $q$ from vanishing requires us to move only through cholesteric textures, analogous to requiring that a biaxial nematic be nowhere uniaxial \cite{Gartland1991numerical}, the ``hard" biaxial phase.

Returning to the interpolated field in (\ref{interp}), we can understand the necessity of $\lambda^{\pm 1}$ defects in the $\bf P$ field in light of the constructed $\phi$ field. At small $r$, where ${\bf P}= \hat z$, the pitch axis does not twist and so  $\phi = q_0 z$ plus an arbitrary uniform shift; the level sets are planar and horizontal. At large $r$, where ${\bf P}= \hat r$, the pitch axis also does not twist and so  $\phi = q_0 r$ plus an arbitrary uniform shift; the levels sets are cylinders concentric about $r=0$. The situation is illustrated schematically in Fig.~\ref{interpfig}b. In the biaxial nematic, the analogous transformation could be accomplished by a uniform ``escape up'' or ``escape down'' of one of the triad vectors with decreasing $r$, and the same rotation of another triad vector with increasing $r$. However, for a cholesteric, such a resolution is incompatible with the existence of a $\phi$ field that maintains nonzero layer spacing (except possibly at line defects). Instead, the layers formed by level sets of $\phi$ must suffer dislocations, realized as disclination dipoles, line defect pairs with $\pm \pi$ winding, as illustrated schematically in Fig.~\ref{interpfig}c. Under our assumptions, these defects in $\boldsymbol{\nabla} \phi$ are also defects in $\bf P$. Since $\bf n$ is nonsingular, the defects are $\lambda^{\pm 1}$ rings. Although the cholesteric configuration might not geometrically admit a foliation everywhere perpendicular to ${\bf P}$, its topology is nonetheless limited by the topologies available to a family of layers. This example illustrates that the algebra of cholesteric defects differs from that of biaxial nematics.

Are all smectic textures {\sl subsets} of the cholesteric textures with the added constraint that ${\bf P}\propto \boldsymbol{\nabla}\phi$?  As long as the pitch axis remains straight, one can make this identification using $\phi$ provided that the distortions are {\sl purel}y $z$-dependent.  Similar identifications are possible for spherically symmetric (hedgehog) and cylindrically symmetric (jelly roll) configurations.  The Volterra construction is then used to create dislocations in the cholesteric in analogy with those in a smectic.  However, is it possible to apply the Volterra construction while preserving the cholesteric structure implied in (\ref{eq:C})?  In general, the answer is no: The topology of a screw defect in a smectic {\sl cannot} be constructed in a cholesteric. This implies that the identification between smectics and cholesterics fails in the context of topological defects. 

To see this, consider the change in angle of the director field along an infinitesimal shift $\epsilon$ along the pitch axis, $
\Delta\theta = -\epsilon\, \nbar\cdot\left(\vec{P}\cdot\boldsymbol{\nabla}\right)\vec{n} = \epsilon q$
where $\Delta\theta$ and $q$ will, in general, depend upon which integral curve of $\vec{P}$ we are considering.  We see that $q=d\theta/d\epsilon$. In this context a Lagrangian set of co\"ordinates is natural -- each point in space is labeled by the arclength along each integral curve, and the integral curve is labeled by its co\"ordinates at one of the two-dimensional boundaries.   However, if the pitch axis is supposed to be perpendicular to the ``smectic layers'' defined by the same value of rotation $\theta$ around each integral curve, then $\vec{P}\propto\boldsymbol{\nabla}\theta$ and we must implicitly switch between Lagrangian co\"ordinates and the Eulerian co\"ordinates in space in order to form the gradient.  Accordingly, $\vec{P}=\boldsymbol{\nabla}\theta/\vert\boldsymbol{\nabla}\theta\vert$ and $q=\pm\vert\boldsymbol{\nabla}\theta\vert$ depending on the sign of twist, held constant throughout the sample as before.   Consider now a standard smectic screw dislocation in $\theta$ with layer spacing $a$ and Burgers scalar $b$, 
level sets of 
$\theta(\vec{x}) = [z - b\arg(x+iy)]/a$
with unit normal $\vec{N}(\vec{x})=[by/r^2,-bx/r^2,1]/\sqrt{1+b^2/r^2}$ where $r=\sqrt{x^2+y^2}$.
One can verify that the unit speed curves
$\vec{R}(s) = \left[ A\cos(\gamma s),-A\sin(\gamma s),B\gamma s\right]
$
satisfy
$\vec{P} = \dot{\vec{R}} =\vec{N}\left[\vec{R}(s)\right]$
where $\gamma= 1/\sqrt{A^2+B^2}$ and $B = A^2/b$.   Finally, as $r\rightarrow\infty$ the torsion of this curve is
$
q=b/({b^2+A^2})
$
while  $\vert\boldsymbol{\nabla}\theta\vert = \sqrt{b^2+A^2}/(A a)$.  We can consider ever larger helical paths by letting $A$ grow, and find that $q\rightarrow 0$ as $A\rightarrow\infty$ -- the integral curves become straight lines.  At the same time, however, $\vert\boldsymbol{\nabla}\theta\vert \rightarrow 1/a$.  These two boundary conditions are incompatible, $q\ne\vert\boldsymbol{\nabla}\theta\vert$, and prevent us from decorating the smectic screw dislocation with a director field.  Can we embellish the integral curves to maintain the torsion?  Recall that there is a family of helices \cite{geometryrmp} that maintain constant torsion while their curvature vanishes.  However, such a family of concentric curves will have single- or double- twist in the core -- a vector field that cannot be foliated since ${\vec N}\cdot(\boldsymbol{\nabla}\times{\bf N})\ne 0$.

To summarize, we have proposed a definition of the pitch axis in the cholesteric that leads to a rich geometry of an orthonormal triad consisting of the pitch axis ${\bf P}$, the director field ${\bf n}$, and the perpendicular direction $\nbar$.    Our construction explicitly demonstrates the differences between the algebra of the topological defects in smectics, cholesterics, and biaxial nematics -- phases with apparently similar or identical ground state manifolds.  The issue is not the ground states or even the states with topological defects; for homotopy theory to apply it must be possible to smoothly distort the director complexion between any two representatives of the same class.  Because of the restrictions on the pitch axis in cholesterics and the layer normal in smectics, algebraic ``equivalence classes'' are {\sl not} homotopic -- the adventure begins.

We acknowledge stimulating discussions with B.G.\ Chen and V.\ Vitelli and support from the UK EPSRC and NSF Grant DMR12-62047.
 D.A.B.\ was supported by NSF Graduate Research Fellowship DGE-1321851. T.M. was supported by a University of Warwick Chancellor's International Scholarship.  S.\v{C}. was partially supported by Slovenian Research Agency contract P1-0099. R.D.K.\ was partially supported by a Simons Investigator grant from the Simons Foundation. R.A.M.\ was supported from FAPESP grant 2013/09357-9.


\begin{thebibliography}{10}
\bibitem{f2}{ Symmetrizing $C_{ij}$ would alleviate this problem but despoil the ensuing geometric description of the triad.}
\end{thebibliography}
\end{document}